\documentclass[10pt]{article}
\usepackage{amssymb,amsmath}
\usepackage{graphicx}
\usepackage{color}
\usepackage{curves,multicol}
\usepackage{geometry}
\usepackage{times}
\usepackage{bm}
\usepackage{ulem}
\usepackage{hyperref}
\usepackage{ucs}
\usepackage[utf8x]{inputenc}
\usepackage[intoc]{nomencl}
\usepackage{float}
\usepackage{subfig}
\usepackage{units}
\usepackage{titlesec}
\usepackage{mathptmx}
\usepackage{lscape}
\usepackage{nopageno}
\usepackage[labelfont=bf]{caption}
\usepackage{lipsum}   
\usepackage[none]{hyphenat} 
\usepackage{breqn}
\makeatletter

\renewcommand{\section}{\@startsection
{section}
{1}
{0mm}
{-\baselineskip}
{0.5\baselineskip}
{\normalfont\bfseries\MakeUppercase}} 

{

\renewcommand{\subsection}{\@startsection
{subsection}
{2}
{0mm}
{0.5\baselineskip}
{0.25\baselineskip}
{\bfseries\normalsize}} 

\makeatother

\newcommand{\Fro}{\text{\textit{Fr}}}
\newcommand{\Rey}{\text{\textit{Re}}}

\newcommand{\Str}{\text{\textit{St}}}

\setlength{\parskip}{0pt}
\setcounter{secnumdepth}{-1} 
\begin{document}
\fussy

\vspace*{-1.0cm}
\begin{flushright} \vbox{
34$^{\mathrm{th}}$ Symposium on Naval Hydrodynamics\\
Washington, DC, USA, 26 June-1 July 2022}
\end{flushright}

\vskip0.65cm
\begin{center}
\textbf{\LARGE
Modal analysis of bluff body wakes\\[0.35cm]
}

\Large S. Nidhan$^{1}$, D. Gola$^{1}$, and S. Sarkar$^{1}$\\ 

($^{1}$University of California San Diego, La Jolla, USA)\\
\vspace*{0.25cm}

\end{center}

\begin{multicols*}{2}

\section{Abstract}

Spectral proper orthogonal decomposition (SPOD) analyses are performed for unstratified and stratified wakes of a circular disk at Reynolds number, $U_\infty D/\nu = 50,000$, where $U_\infty$ is the freestream velocity and $D$ is the diameter of the circular disk. At $\Rey = 50,000$, three diameter-based Froude numbers, $\Fro = U_\infty/ND = \infty$, $2$ and $10$ are analyzed. We find that in the unstratified configuration, two modes: (i) the vortex shedding (VS) mode ($m=1, \Str = 0.135$) and (ii) the double helix (DH) mode ($m=2, \Str \rightarrow 0$) are the most dominant features. At far-downstream distances, DH mode is dominant while at near and intermediate locations, VS mode dominates the energetics. Reconstruction of TKE and Reynolds shear stresses are also explored at near and far wake locations. In the stratified cases ($\Fro = 2$ and $10$), the vortex shedding mode (at $\Str \approx 0.13-0.14$) dominates the wake. At intermediate to late streamwise locations ($x/D$), we establish through pressure wave flux reconstruction that the VS mode is responsible for the internal gravity wave (IGW) generation. A comparison between the SPOD decomposition analyses between $\Rey  =5000$ and $\Rey = 50,000$ is also performed at $\Fro =2$. Preliminary results show that the low-$\Rey$ wake is also dominated by the VS mode but exhibits more coherence, namely higher contribution of leading SPOD modes, than the high-$\Rey$ wake.
 
\section{Introduction}
\label{SECintroduction}

Townsend \cite{townsend1976structure} hypothesized that turbulent wakes forget their initial conditions far from the wake generator. However experimental \cite{bevilaqua_turbulence_1978} and numerical \cite{redford_universality_2012,ortiz-tarin_high-reynolds-number_2021} studies have shown that the effect of initial conditions can remain embedded in these flows for large downstream distances ($x/D$). Hence, it is expected that initial conditions, like the shape and characteristic of wake generator, can significantly influence the coherent structures \cite{taneda1978visual,cannon1993observations} in the wake.  These structures can carry significant amount of turbulent kinetic energy (TKE) and Reynolds stresses. Hence, an in-depth understanding of coherent structures is fundamental to uncover the complete picture of how a turbulent wake evolves.

In the presence of stratification, several distinctive features emerge in turbulent wakes, e.g. steady lee-waves in the downstream \cite{ortiz-tarin_stratified_2019}, multistage wake decay \cite{spedding_evolution_1997}, unsteady internal gravity waves (IGWs) at intermediate to late buoyancy times $Nt(=(x/D) (1/Fr))$ \cite{abdilghanie_internal_2013,rowe_internal_2020} and emergence of pancake vortices \cite{spedding_streamwise_2002}. Previous works of Bonneton et al. \cite{bonneton_internal_1993} and Lin et al. \cite{lin_turbulent_1992} have shown the existence of vortex shedding (VS) frequency at select locations in the near wake at intermediate Reynolds numbers. However, there are still some open questions: (i) Is the VS mode present even at large downstream locations in stratified wakes? (ii) How energetic are the coherent structures in stratified wakes compared to the unstratified counterpart?, and (iii) How much do they contribute to the unsteady IGW generation mechanism? 

To this end, the objective of this work was to analyze the energetics of coherent structures in unstratified and stratified wakes at a moderately high  $\Rey$. We use the datasets of Chongsiripinyo and Sarkar \cite{chongsiripinyo_decay_2020} of flow past a circular disk at $\Rey = U_\infty D/\nu = 50,000$ and $\Fro = U_\infty/ND = \infty, 2, 10$. Here, $U_\infty$, $D$, $N$, and $\nu$ correspond to free-stream velocity, body diameter, buoyancy frequency and kinematic viscosity of the fluid. Spectral proper orthogonal decomposition (SPOD) is used to analyze the evolution of these coherent structures for downstream location extending to $x \sim O(100D)$. SPOD has been shown to effectively distill the coherent features in various different flow configurations, e.g.  wall bounded flows \cite{muralidhar_spatio-temporal_2019,abreu2020spectral}, turbulent jets \cite{schmidt_spectral_2018}, etc. Contrary to methods like dynamic mode decomposition \cite{nidhan_dynamic_2019}, SPOD provides an ordered set of modes at a given frequency $\Str$. In this paper, we summarize the key findings of Nidhan et al. \cite{nidhan_spectral_2020,nidhan2022analysis} for the $\Rey = 50,000$ wake. In addition, we present preliminary comparison between SPOD analyses of flow past a circular disk at $\Rey = 5000$, $\Fro= 2$ with $\Rey = 50,000$ and $\Fro = 2$ results. The aim of this comparison is to assess the change in modal energy content when $\Rey$ is changed. In future, we wish to leverage the findings from this ongoing study to provide realistic inflow conditions to body-exclusive simulations \cite{vandine_hybrid_2018} for high $\Rey$ wakes.

\section{Methodology}
\label{SECmethods}

For the numerical implementation of SPOD, two-dimensional data (at a given $x/D$) of all flow variables at each timestep is flattened into a column vector and assembled in a snapshot matrix $\mathbf{P} = [\mathbf{q}^{(1)},  \mathbf{q}^{(2)}, \cdots, \mathbf{q}^{(N)}]$, where $\mathbf{q}^{(i)} = [\mathbf{u}^{(i)}, \rho^{(i)}]^{T}$ is the $i^{th}$ temporal snapshot  and $N$ is the total number of snapshots employed for the SPOD. Thereafter, the temporal mean $\bar{\mathbf{q}} = \frac{1}{N}\sum^{N}_{i=1}\mathbf{q}^{(i)}$ is subtracted from the snapshot matrix:

\begin{equation}
\mathbf{Q} = [\mathbf{q}^{(1)}-\bar{\mathbf{q}},  \mathbf{q}^{(2)}-\bar{\mathbf{q}}, \cdots, \mathbf{q}^{(N)}-\bar{\mathbf{q}}] = [\mathbf{q'}^{(1)},  \mathbf{q'}^{(2)}, \cdots, \mathbf{q'}^{(N)}].
\label{turb_matrix}
\end{equation}

The $\mathbf{Q}$ matrix is then divided into $N_{blk}$ overlapping blocks with $N_{freq}$ entries in each block as follows:
\begin{equation}
\mathbf{Q}^{(l)} = [\mathbf{q'}^{(l)(1)},  \mathbf{q'}^{(l)(2)}, \cdots, \mathbf{q'}^{(l)(N_{freq})}],
\label{block_matrix}
\end{equation}
where $l$ denotes the $l^{th}$ block of data containing $N_{freq}$ temporal snapshots. Each $\mathbf{Q}^{(l)}$ block is then Fourier transformed in time resulting in the $\hat{\mathbf{Q}}$ matrix: 
\begin{equation}
\hat{\mathbf{Q}}^{(l)} = [\hat{\mathbf{q}}^{(l)(1)},  \hat{\mathbf{q}}^{(l)(2)}, \cdots, \hat{\mathbf{q}}^{(l)(N_{freq})}],
\end{equation}
Following this, all the Fourier realization at a specific frequency $f$ from $N_{blk}$ matrices are collected into a a single matrix $\hat{\mathbf{Q}}_{f}$ as follows:
\begin{equation}
\hat{\mathbf{Q}}_{f} = [\hat{\mathbf{q}}^{(1)(f)},  \hat{\mathbf{q}}^{(2)(f)}, \cdots, \hat{\mathbf{q}}^{(N_{blk})(f)}].
\end{equation}
Once $\hat{\mathbf{Q}}_{f}$ is formed, we obtain the SPOD eigenvalues and eigenmodes by eigenvalue decomposition of the weighted cross-spectral density matrix $\hat{\mathbf{Q}}^{*}_{f}\mathbf{W}\hat{\mathbf{Q}}_{f} \boldsymbol{\Gamma}_{f} = \boldsymbol{\Gamma}_{f}\boldsymbol{\Lambda}_{f}$ where $\boldsymbol{\Lambda}_{f} = \mathrm{diag}\Big(\lambda^{(1)}_{f},\lambda^{(1)}_{f}, \cdots \lambda^{(N_{blk})}_{f}\Big)$ is a diagonal matrix containing $N_{blk}$ eigenvalues ranked in  decreasing order of energy content from $i = 1$ to $N_{blk}$. The corresponding spatial eigenmodes $\hat{\boldsymbol{\Phi}}_{f}$ can be obtained as $\hat{\boldsymbol{\Phi}}_{f} = \hat{\mathbf{Q}}_{f}\boldsymbol{\Gamma}_{f}\boldsymbol{\Lambda}_{f}^{-1/2}$. $\mathbf{W}$ is a diagonal matrix containing the numerical quadrature weights. Multiplication by $\mathbf{W}$ ensures that the obtained SPOD eigenvalues optimally capture the area-integrated sum of the defined energy norm at any given cross-section. 

Since the $\Fro = \infty$ wake has azimuthal symmetry, its snapshots are first azimuthally decomposed and then eigenvalue decomposition is performed for each azimuthal wavenumber $m$. Note that the azimuthal decomposition is not performed for the stratified cases as the stratification kills the azimuthal symmetry in the flow. For all cases, we perform SPOD one axial location ($x$) at a time due to significant computational cost of performing a $3D$ SPOD in wakes that cover downstream location until $x/D = 125$. For $3D$ SPOD, a single snapshot entry would have comprised of $O(10^9)$ elements, making the computation intractable for $O(10^4)$ snapshots.

The parameters for SPOD analysis  of stratified cases at $\Rey = 50,000$ are as follows: (i) total number of snapshots $N=7168$ with consecutive snapshots separated by $\Delta tD/U_{\infty} = 0.09$ and $0.104$ for $\Fro = 2$ and $10$, respectively, (ii) number of frequencies $N_{freq} = 512$, and (iii) overlap between blocks $N_{ovlp} = 256$, resulting in total of $N_{blk} = \frac{N-N_{ovlp}}{N_{freq} - N_{ovlp}} = 27$ SPOD modes at each frequency. For the  $\Fro = \infty$ wake, $7200$ snapshots with $\Delta tD/U_{\infty} = 0.07$ are employed while keeping the other SPOD parameters same as in the stratified cases. For $\Rey = 5000$ at $\Fro = 2$, $N_{freq}$ and $N_{ovlp}$ are kept the same with $\Delta t D/ U_{\infty} = 0.07$ and $N = 5429$ which gives $N_{blk} = 20$ SPOD modes for this case. Readers are referred to Towne et al. \cite{towne_spectral_2018} and Schmidt and Colonius \cite{schmidt2020guide} for more details regarding the theoretical and numerical aspects of SPOD algorithm.
 
\section{Results}
\label{SECresults}

\subsection{SPOD analysis of unstratified wake at $\Rey = 50,000$}
\label{SECFrinf}
\noindent

\begin{figure}[H]
	\includegraphics[width=\linewidth]{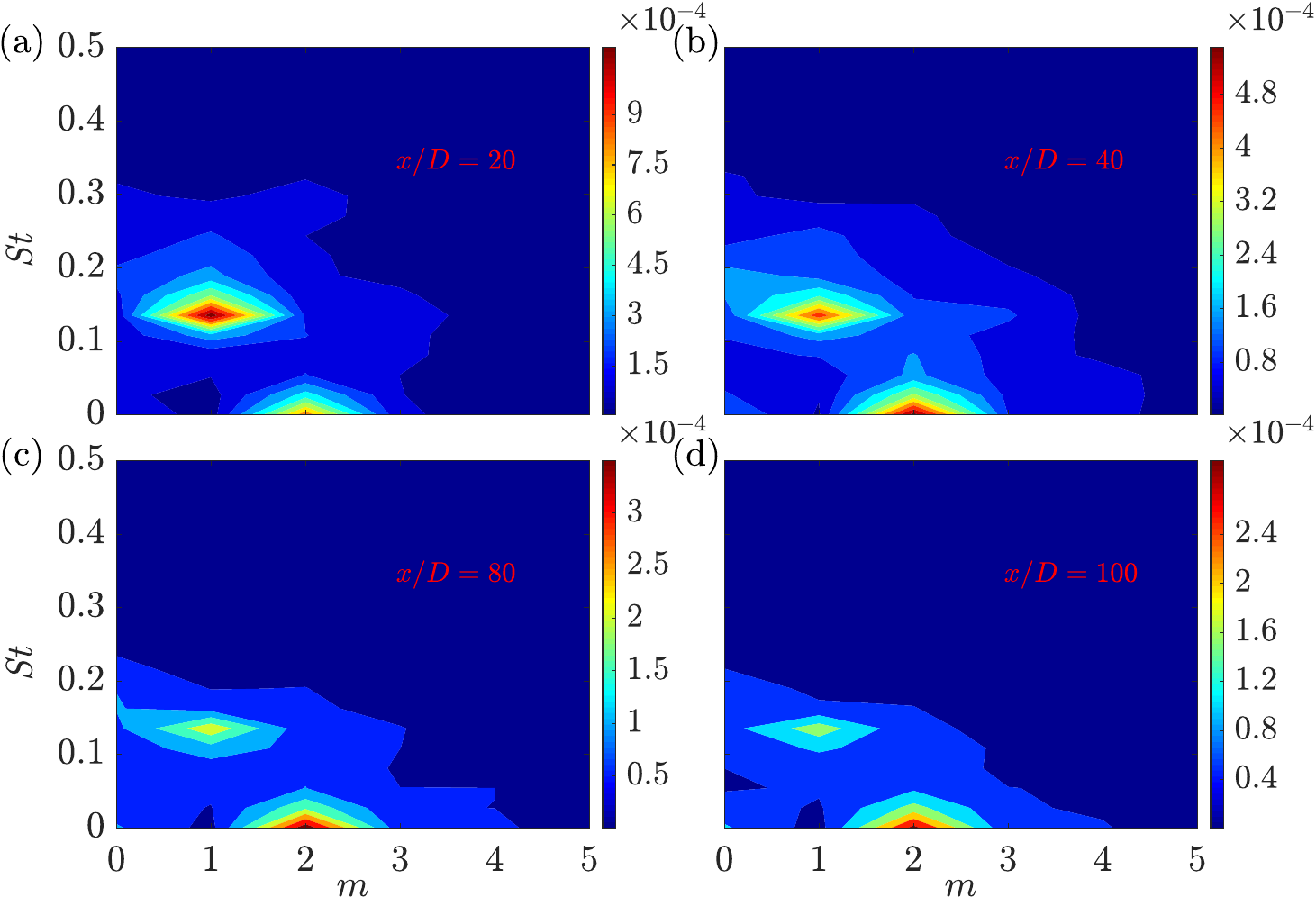}
	\caption{SPOD contour maps showing energy contained in leading SPOD mode, $\lambda^{(1)}$ as a function of ($m$, $\Str$) at $x/D = 20, 40, 80$ and $100$. The color bar limits are set according to the maximum values of $\lambda^{(1)}$ over all ($m$, $\Str$) pairs at the respective $x/D$ locations.}
	\label{fig:contour_maps_eigenvalues}  
\end{figure}

Energetics in the unstratified wake is found to be dominated by two distinct coherent structures at intermediate and large downstream locations as shown in figure \ref{fig:contour_maps_eigenvalues}: (i) the vortex shedding (VS) mode with $m=1, \Str =0.135, n=1$, and (ii) the double helix (DH) mode with $m=2, \Str \rightarrow 0, n=1$. Here, $m$ is the azimuthal mode number, $n$ is the SPOD modal index, and $\Str = fD/U_\infty$ is the Strouhal number. 

While the $m=1, \Str = 0.135$ has long been attributed to the vortex shedding mechanism in the wake of a disk \cite{fuchs_large-scale_1979,berger_coherent_1990, cannon1993observations}, the importance of the DH mode was first elucidated by Johansson et al. \cite{johansson_far_2006-1}. The prominent peak at $\Str=0$ should be interpreted as a quasi-steady structure in the limit of $\Str \rightarrow 0$ given the discrete nature of the Fourier transform employed for the numerical data. From figure \ref{fig:contour_maps_eigenvalues}, one can also infer that the dominance of the VS mode in the intermediate wake gives way to the dominance of the DH mode at late downstream locations. Additionally, we find that the overall dominance of the azimuthal mode $m=2$ over $m=1$ appeared around $x/D \approx 60$. At the same streamwise location,  the mean defect decay also transitions from the non-equilibrium decay  \cite{vassilicos_dissipation_2015} of $U_d \sim x^{-0.9}$ to the classical decay of $\sim x^{-2/3}$ \cite{chongsiripinyo_decay_2020}. 

\begin{figure}[H]
	\includegraphics[width=8cm]{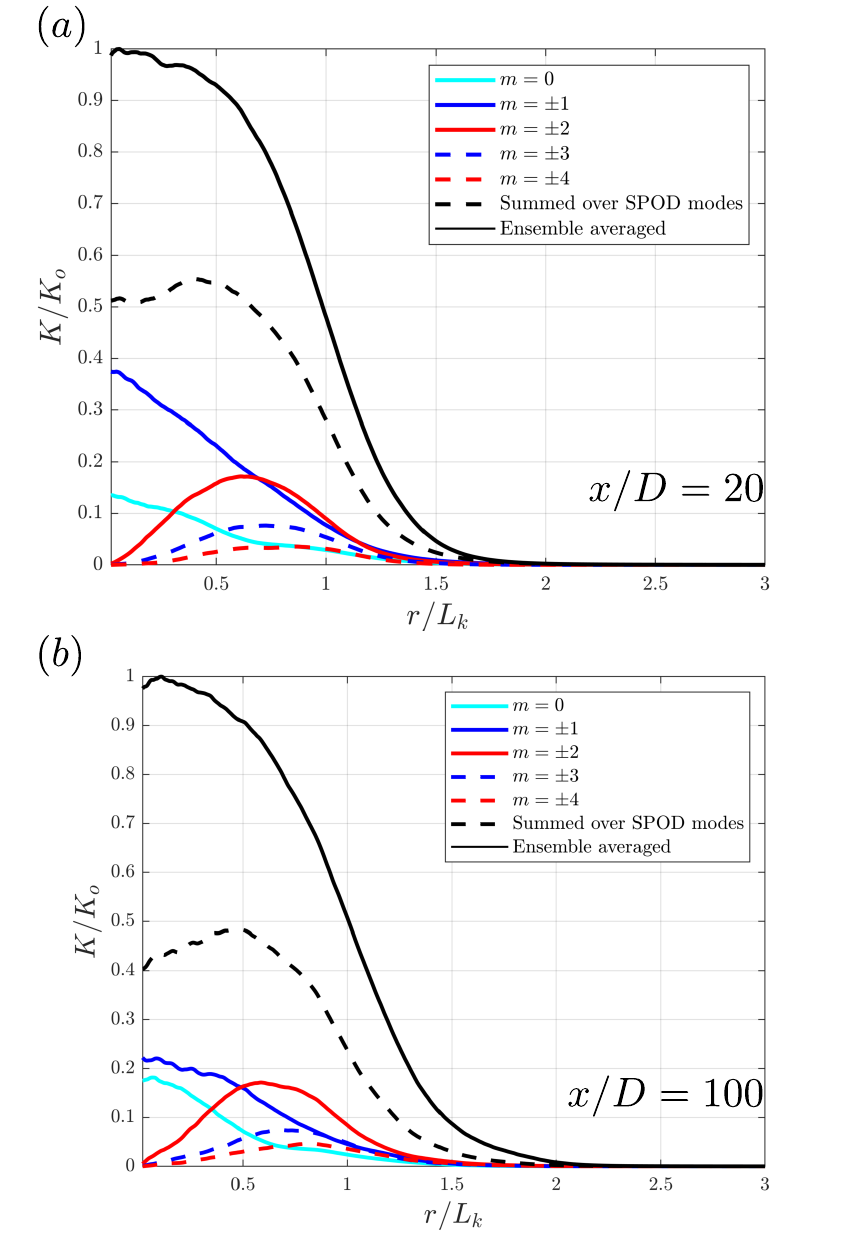}
	\caption{Reconstruction of TKE from a low-order truncation that comprises the leading 3 SPOD modes of $-4 \leq m \leq 4$ with energy summed over $-1 \leq\Str \leq 1$ at (a) $x/D = 20$ and (b) $x/D = 100$. The radial direction is scaled with TKE-based wake width $L_{k}$ and TKE is scaled with its centerline value, $K_{o}$.}
	\label{fig:recon_tke}  
\end{figure}

The reconstruction quality of TKE and $\langle u'_{x}u'_{r} \rangle$ using SPOD modes has also been assessed. By definition, SPOD modes are constructed to be optimal for the area-integrated TKE and an accurate reconstruction of $\langle u'_{x}u'_{r} \rangle$ is not guaranteed. However, we find that the fidelity of SPOD modes in reconstructing the Reynolds shear stress is significantly better compared to the TKE reconstruction. 

Figure \ref{fig:recon_tke} shows the TKE reconstruction using SPOD modes at two downstream locations of $x/D = 20$ and $100$ using the ($m,\Str,n$) triplets mentioned in the caption.  At  the centerline, the qualitative trend of the reconstructed TKE remains similar with approximately $40-50\%$ accuracy at both intermediate and far wake locations. Away from the wake core, the reconstruction quality improves owing to  the less turbulent state of the flow away from the centerline. It can also be noted that the reconstruction quality slightly deteriorates in the centerline with increasing $x/D$. $m=0-2$ are the leading individual azimuthal contributors to the overall TKE.

\begin{figure}[H]
	\includegraphics[width=8cm]{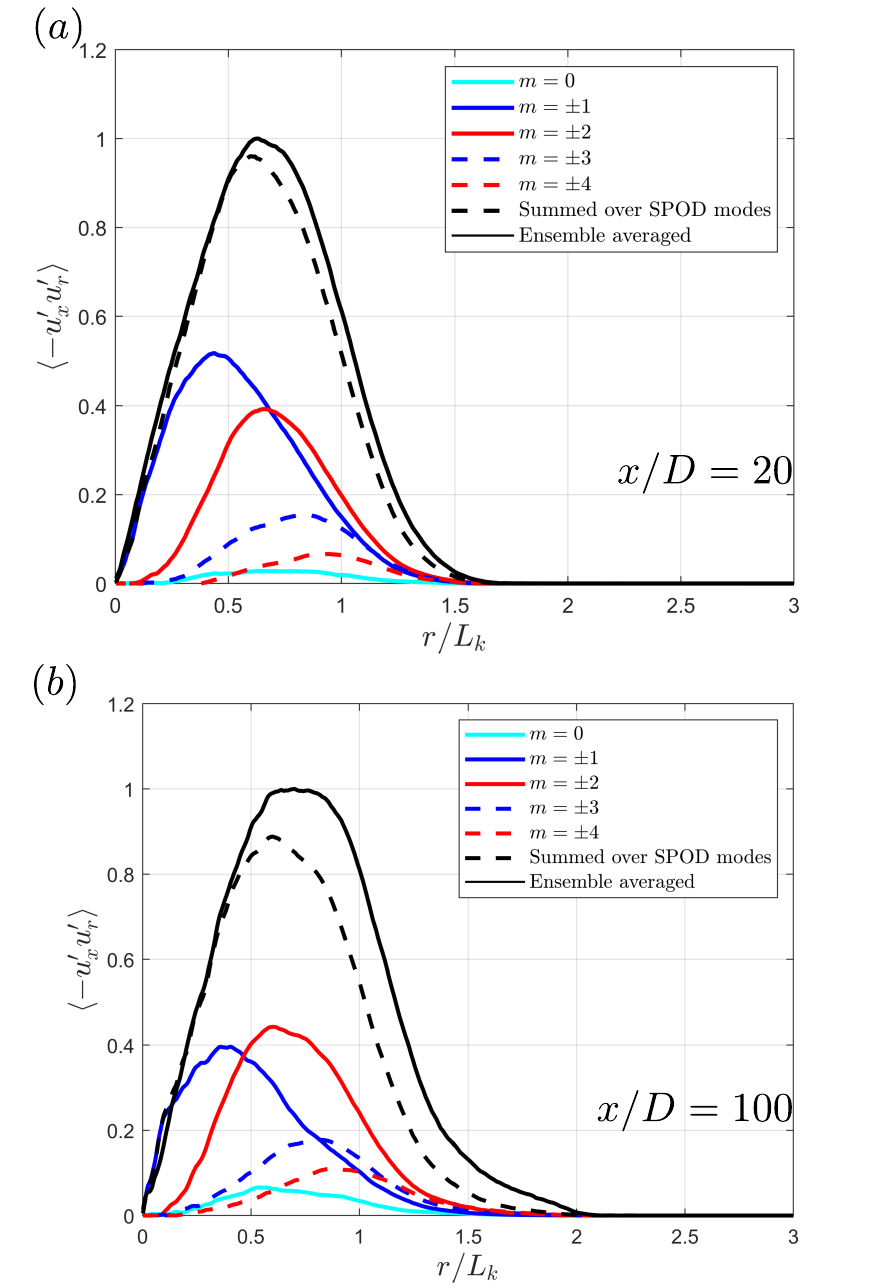}
	\caption{Reconstruction of $\langle u'_{x}u'_{r} \rangle$ from the leading 3 SPOD modes of various $-4 \leq m \leq 4$ summed over $-1 \leq\Str \leq 1$ at (a) $x/D = 20$ and (b) $x/D = 100$. Radial direction is scaled with TKE-based wake width $L_{k}$ and $\langle u'_{x}u'_{r} \rangle$ is scaled with $\langle -u'_{x}u'_{r} \rangle_{max}$.}
	\label{fig:recon_uxur}  
\end{figure}

Likewise, figure \ref{fig:recon_uxur} compares the reconstruction of $\langle u'_{x}u'_{r} \rangle$ with the corresponding actual value at four locations $x/D = 20$ and $100$. The same sets of modes as in figure \ref{fig:recon_tke} are employed for reconstruction. The reconstruction turns out to have significantly higher fidelity for  $\langle u'_{x}u'_{r} \rangle$ than for TKE. Azimuthal modes $m=1$ and $m=2$ are the two major contributors to the Reynolds shear stress followed by $m=3$ and $4$. $m=0$ mode carries significantly less $\langle u'_xu'_r \rangle$ in contrast to its high contribution towards TKE. Interested readers are referred to Nidhan et al. \cite{nidhan_spectral_2020} for an in-depth analysis on trends of TKE and $\langle u'_xu'_r \rangle$ reconstruction.

\subsection{SPOD analysis of stratified wakes at $\Rey = 50,000$}
\label{SECFr2Fr10}
\begin{figure}[H]
	\includegraphics[width=0.9\linewidth]{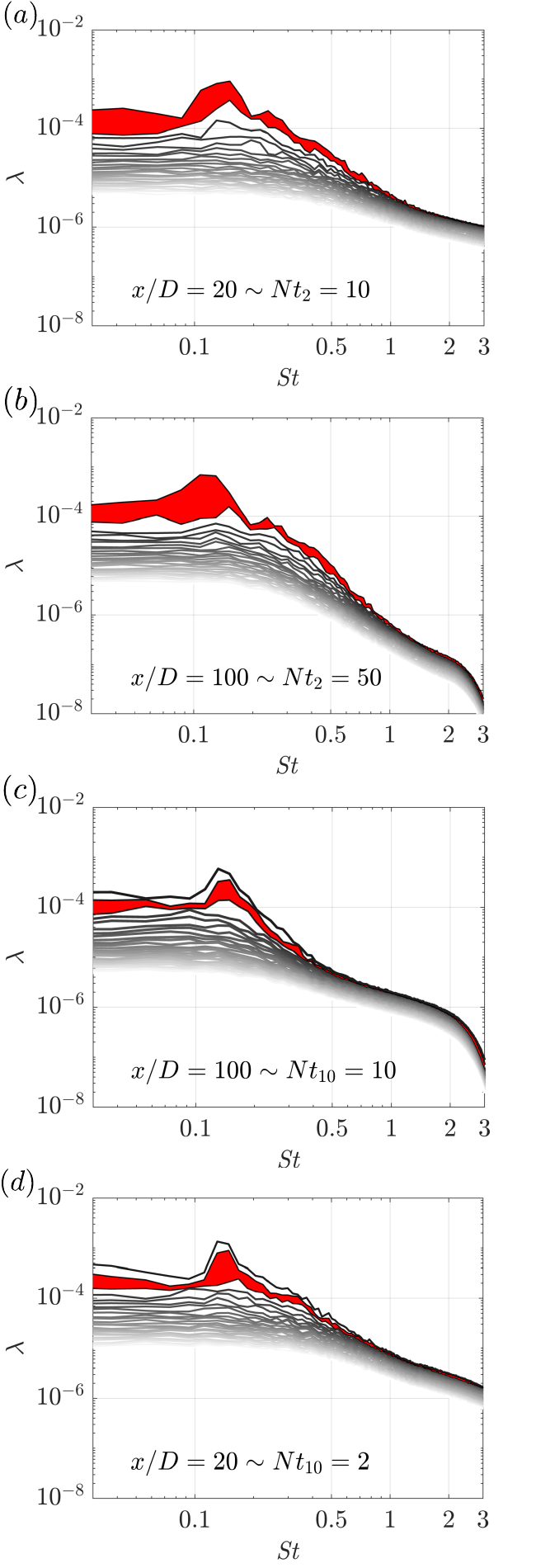}           
	\caption{Eigenspectra of $\Fro = 2$ (a,b) and $\Fro = 10$ (c,d) wakes at $x/D = 20$ and $100$. Leading 25 eigenvalues are shown. Dark to light shade corresponds to SPOD eigenvalues in decreasing order of energy content.}
	\label{fig:spod_spectra_fr2fr10}  
\end{figure}

\noindent
Figure \ref{fig:spod_spectra_fr2fr10} shows the SPOD eigenspectra in the intermediate ($x/D = 20$) and far wake ($x/D = 100$) for $\Fro = 2$ and $10$. In all four plots, a prominent peak at $\Str \approx 0.13-0.14$ can be observed. As discussed in the previous section, this frequency corresponds to the vortex shedding (VS) frequency in flow past a circular disk in unstratified environment. There is a significant gap between the optimal and sub-optimal SPOD modes around VS frequency (shaded by red). This implies that the dynamics in the proximity of the VS mode is low-rank, being primarily dominated by leading SPOD modes at the VS frequency. At $\Str \gtrapprox 0.5$, energy in all modes steeply decreases in both wakes. In the $\Fro=10$ wake, we shade the region between second and third eigenvalues since the first two SPOD modes are similar in energy content. On further investigation, we find that the first and second SPOD eigenmodes at the VS frequency in the $\Fro=10$ wake have a similar spatial structure, but with a rotation in their orientation. We hypothesize that these two modes at the VS frequency are the manifestation of $m = \pm 1$ azimuthal modes in the weakly stratified $\Fro =10$ wake.

\begin{figure}[H]
	\includegraphics[width=0.82\linewidth]{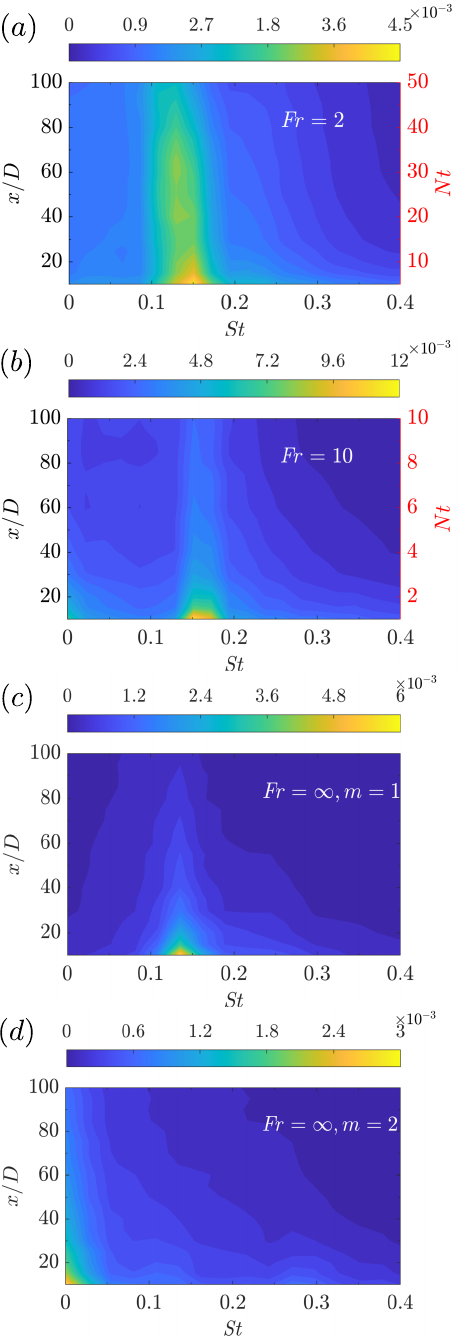}
	\caption{$x/D-\Str$ variation of total energy in leading 15 SPOD modes for (a) $\Fro = 2$, (b) $\Fro = 10$, (c) $\Fro = \infty, m=1$, and (d) $\Fro = \infty, m = 2$.}
	\label{fig:fr2fr10_active_freq}  
\end{figure}

Figure \ref{fig:spod_spectra_fr2fr10} establishes that the VS mechanism is active in the stratified wakes even at far downstream location ($x/D = 100$). To compare how energetic is the VS mode in stratified wakes compared to dominant modes in the unstratified wake, we compare the $x/D-\Str$ variation of the energy contained in the leading 15 SPOD modes across $\Fro = \infty$, $2$ and $10$ wakes in figure \ref{fig:fr2fr10_active_freq}. For $\Fro = \infty$, variation of energy in $m=1$ and $m=2$ modes, the two most energetic azimuthal modes, are shown. Figure \ref{fig:fr2fr10_active_freq} shows that the decay of energy around the dominant frequencies is the slowest for the $\Fro = 2$, followed by $\Fro = 10$ and $\Fro = \infty$ wakes, respectively. This observation establishes that the body-induced coherence is sustained for long distances in the stratified wakes as compared to their unstratified counterpart. 


\begin{figure}[H]
	\includegraphics[width=0.82\linewidth]{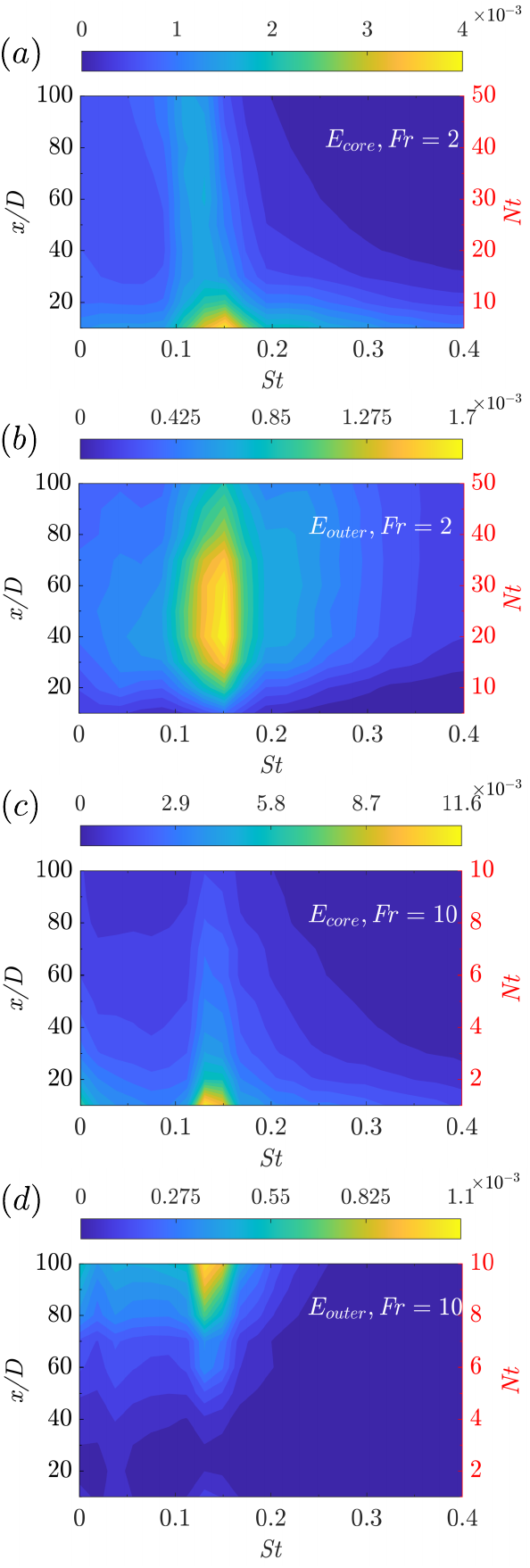}
	\caption{$x/D-\Str$ variation of total energy in leading 15 SPOD modes partitioned between wake core (a,c) and outer wake (b,d) for $\Fro =2$ and $\Fro = 10$ wakes.}
	\label{fig:fr2fr10_parition}  
\end{figure}
\begin{figure}[H]
	\includegraphics[width=1\linewidth]{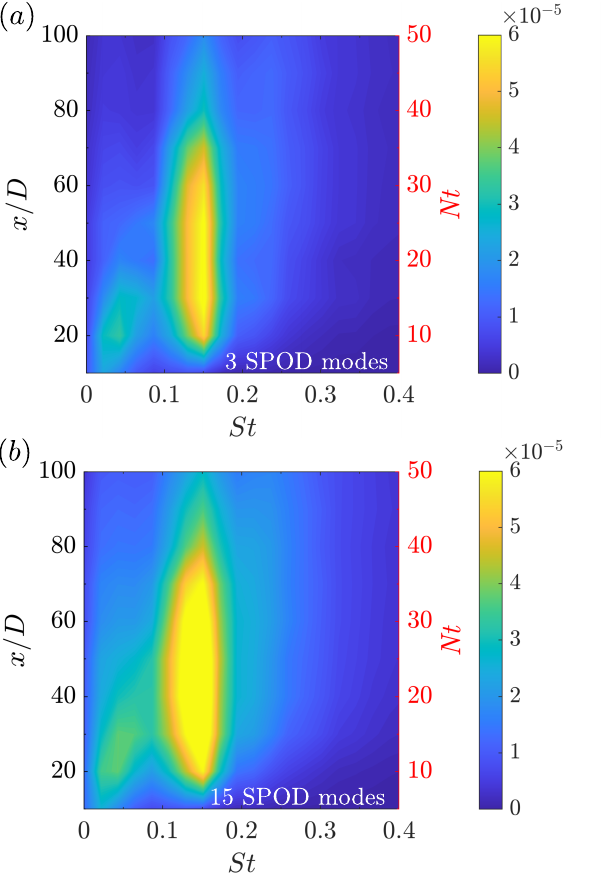}
	\caption{$x/D-\Str$ contour maps showing the variation of $\langle p'u'_r\rangle$, integrated in the outer wake region for $\Fro = 2$ wake, reconstructed from (a) 3 SPOD modes and (b) 15 SPOD modes.}
	\label{fig:pur_reconstruction}  
\end{figure}

To investigate the reason behind the sustenance of the coherent features in the stratified wakes for large $x/D$, we plotted the energy partitioned between wake core and outer wake region for $\Fro = 2$ and $\Fro = 10$ wakes as shown in figure \ref{fig:fr2fr10_parition}. The wake core is defined by an ellipse with major and minor axes being two-times horizontal TKE based wake width (Chongsiripinyo and Sarkar \cite{chongsiripinyo_decay_2020}) and vertical TKE based wake width at a given $x/D$. The energy in the wake core of both $\Fro$ (figure \ref{fig:fr2fr10_parition}(a,c)) decay with $x/D$. However, in the outer wake, we see accumulation of energy in (a) $\Fro = 2$ wake from $20 < x/D < 80$ ($10<Nt<40$) and (b) $\Fro = 10$ wake from $60 < x/D < 100$ ($6 < Nt < 10$). The energy in outer wake is concentrated in the vicinity of the VS frequency. At similar $x/D$ locations, visualization of SPOD modes at the VS frequency (not shown here for brevity), show IGWs emanating from wake core to outer wake. Other numerical works employing temporal model for wakes, e.g., deStadler and Sarkar \cite{de_stadler_large_2014}, Abdilghanie and Diamessis  \cite{abdilghanie_internal_2013} and Rowe et al. \cite{rowe_internal_2020} have also found emission of IGWs in the wake core for $20 < Nt < 70$. Results from our SPOD analyses are in qualitative agreement with these work. Going a step further, SPOD analysis establishes that these IGWs are primarily emitted at the VS frequency (at least in wakes with $\Fro > 2$), indicating a strong causal link between the VS mechanism and IGW generation in stratified wakes.

Figure \ref{fig:pur_reconstruction} shows the reconstructed radial pressure wave flux, $\langle p'u'_r\rangle$, integrated in the outer wake region for the $\Fro =2$ wake. It is important to note that $\langle p'u'_r\rangle$ is responsible for transferring energy from the wake core to the outer wake through IGWs (Rowe et al. \cite{rowe_internal_2020}). The reconstruction has been performed using leading 3 and leading 15 SPOD modes in figure \ref{fig:pur_reconstruction}(a,b), respectively. Both plots show a strong contribution of the VS mechanism to the pressure wave flux term. The evolution of $\langle p'u'_r \rangle$ with $x/D$ is very similar to the variation of energy in the outer wake region in the $\Fro = 2$ wake (figure \ref{fig:fr2fr10_parition}b). Figure \ref{fig:pur_reconstruction} establishes a causal link between the IGW generation process and the VS mechanism showing that it is the VS mode that contributes primarily to the emission of IGWs.

\subsection{Comparison of the modal content between $\Rey = 5000$ and $\Rey = 50,000$ wakes at $\Fro=2$}
\label{SECRe5k}
\noindent

\begin{figure}[H]
	\includegraphics[width=7cm]{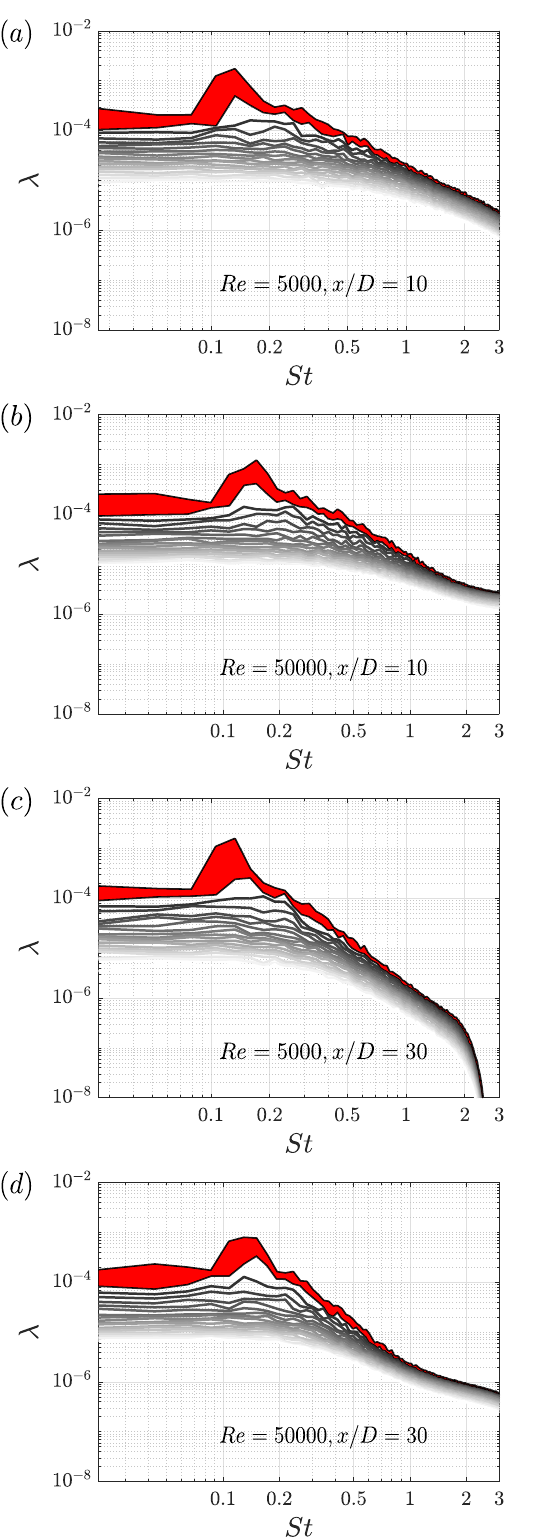}
	\caption{Eigenspectra for $\Fro = 2$ at $\Rey = 5000$ and $50,000$ for $x/D = 10$ and $30$. Leading $20$ eigenvalues are shown. }
	\label{fig:eigspeccomp}  
\end{figure}

We now briefly describe the differences between the modal content at $\Rey = 5000$ and $\Rey = 50,000$. Figure \ref{fig:eigspeccomp} compares SPOD eigenspectra between $\Rey = 5000$ and $\Rey = 50,000$, at $\Fro=2$. Both cases exhibit peaks at the VS frequency of $\Str \approx 0.13-0.14$. However, the two Reynolds numbers differ in the energy difference between the leading two modes. This difference is higher at the lower Reynolds number of $5000$ and appears to increase with increasing $x/D$ or $Nt$. 

\begin{figure}[H]
	\includegraphics[width=0.9\linewidth]{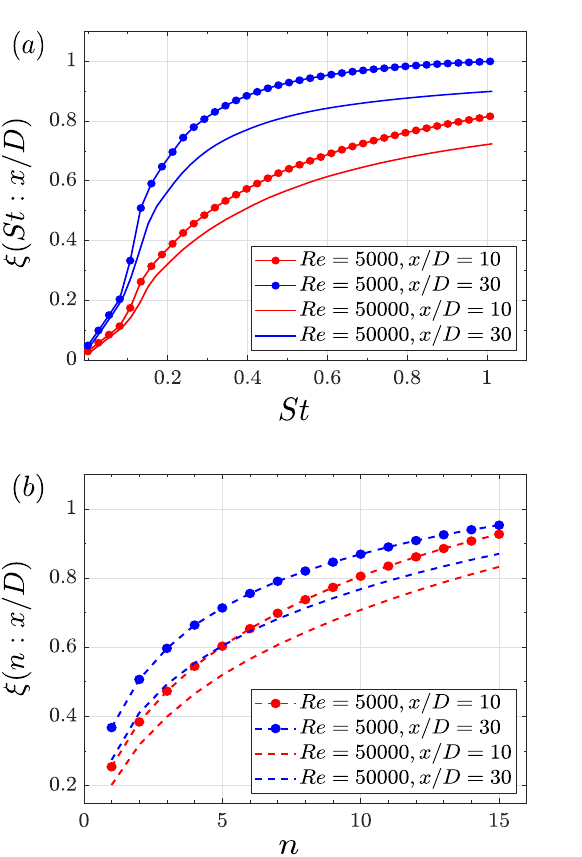}
	\caption{Cumulative energy as a function of frequency (a) and modal index (b).}
	\label{fig:energyvariation}  
\end{figure}

Figure \ref{fig:energyvariation} (a) shows the cumulative energy as a function of the frequency. This energy is calculated by summing over all the $N_{blk}$ number of modes for both the cases and plotting it against frequency. For $\Str < 0.2$ (which includes the VS frequency), there is little difference between the two cases. As $\Str$ increases, the case  with the higher $\Rey$ of $50,000$  has progressively larger value of $\xi(\Str)$  relative to $Re = 5000$, signifying increasing energetic importance of fine-scale motions. It is worth noting that, at $x/D=30$ for $\Rey = 5000$, almost all of the energy is contained in frequencies corresponding to $\Str < 1$. Quantitatively, at $x/D=10$ more than $80 \%$ of the energy in $\Rey = 5000$ is captured in frequencies corresponding to $\Str < 0.5$ and at $x/D = 30$ almost all the energy is captured at frequencies corresponding to $\Str < 1$. A similar comparison is done in figure \ref{fig:energyvariation} (b), where the cumulative energy, now summed over all the frequencies, is plotted against the modal index. We see that the $\Rey = 5000$ case has higher energy content in lower SPOD modes as compared to the $\Rey = 50,000$ cases. The modal indices up to $n = 10$ collectively, are enough to capture more than $80 \%$ of the energy until $x/D = 30$, whereas for $\Rey = 50,000$ that number is $15$.

\section{Conclusions}
\label{SECconclusion}

In this work, we employ spectral proper orthogonal decomposition (SPOD) to analyze the coherent structures in the turbulent wake of a circular disk at $\Rey = 50,000$ and $\Fro = \infty, 2$ and $10$. The analysis encompasses an unprecedented downstream distance of $x/D = 100$ where $D$ is the body diameter and $x$ is the streamwise coordinate. The obtained SPOD eigenvalues are a function of distance ($x/D$) and frequency ($\Str$) in stratified wakes. In the case of unstratified wake, these eigenvalues vary with $x/D$, $\Str$, and $m$ (azimuthal wavenumber). A preliminary comparison between SPOD results of $\Rey = 5000$ and $\Rey = 50000$ is also performed at $\Fro = 2$. In future, we wish to leverage the similarity in the modal content across different $\Rey$ to widen the scope of body-exclusive simulations (VanDine et al. \cite{vandine_hybrid_2018}) in modeling high $\Rey$ flows.

In the $\Fro = \infty$ wake, two modes: (i) the vortex shedding (VS) mode and (ii) the double helix (DH) mode dominate the energetics until $x/D = 100$. For a circular disk, VS mode resides at $m=1$ and $\Str= 0.135$ while the DH mode resides at $m=2, \Str = 0$. At near and intermediate distances, the VS mode is more dominant compared to the DH mode while at farther downstream locations ($x/D > 40$), DH mode emerges as the dominant structure in the flow. This finding is in accord with previous experimental study of Johansson et al. \cite{johansson_proper_2002}. However, additionally we find that the overall dominance of the $m=2$ mode with respect to the $m=1$ mode occurs at the same $x/D$ \cite{nidhan_spectral_2020} where the defect velocity ($U_d$) decay rate changes (Chongsiripinyo and Sarkar \cite{chongsiripinyo_decay_2020}) from $x^{-0.9}$ to $x^{-2/3}$. Besides $m=1$ and $2$ modes, summed contribution from $m=0,3$ and $4$ is also significant. We discuss the evolution of their eigenspectra in more detail in Nidhan et al. \cite{nidhan_spectral_2020}. Reconstruction of second-order statistics such as TKE and Reynolds stress ($\langle u'_xu'_r\rangle$) using SPOD modes further uncover the dominance of $m=1$ and $2$ modes compared to other azimuthal modes at all $x/D$. Interestingly, $\langle u'_xu'_r\rangle$ shows better reconstruction quality than TKE for the same set of modes, pointing towards the low-rank nature of $\langle u'_xu'_r\rangle$.  

SPOD analyses of the stratified wakes reveal the dominance of the VS mechanism in the near and far wake of both $\Fro = 2$ and $10$. Further analyses reveal that the energy in the frequencies around VS decays the slowest for $\Fro = 2$, followed by the $\Fro = 10$ and $\Fro = \infty$ wake, respectively. This implies that the body-induced coherence is prolonged in the stratified wakes compared  to their unstratified counterpart. Partitioning the energy at $x/D=10-100$ between the wake core and outer wake region reveal that the energy appears in the outer wake region for (i) $Nt = 10-40$ in the $\Fro =2$ wake and (ii) $Nt=6-10$ in the $\Fro = 10$ wake. The elevated energy levels in the outer wake region are concentrated around the VS frequency in both wakes, indicating a link between the IGW generation mechanism and VS mechanism. The causal link is further confirmed by reconstructing the radial pressure wave flux $\langle p'u'_r\rangle$ in the $\Fro = 2$ wake using SPOD modes. Reconstructed $\langle p'u'_r\rangle$ shows a strong imprint of the VS mode at similar $Nt$ where we observe increased energy content in the outer wake region of the $\Fro =2$ wake. 


Comparison of the SPOD eigenspectrum between $\Rey = 5000$ and $\Rey = 50,000$ shows that although both the spectra look qualitatively similar with dominant peaks near the disk vortex shedding frequency, the relative contributions of the higher SPOD modes and frequencies are not the same. When comparing the cumulative modal energy content, the results show that fewer SPOD modes and frequencies are required to capture the same amount of energy in the $\Rey = 5000$ wake than in the $\Rey = 50,000$ wake.

Overall, our results show that the shape of wake generator can have a lasting impact on the flow evolution, in both unstratified and stratified conditions. Particularly, in stratified wakes, body-generated VS coherence can dictate the course of IGW generation at intermediate $Nt$. Thus, moving ahead, it is crucial to embed the characteristics of wake generator in the initial conditions to capture accurate flow physics. At higher $\Rey$, explicitly modeling the wake generator becomes prohibitively expensive. Thus, there is a need for alternate approaches to incorporate wake generator features in numerical simulations at higher $\Rey$. Similarity in the large scale coherent structures across lower and higher $\Rey$ can be exploited towards this purpose in the future works.

\section{Acknowledgments}
\label{SECacknowledgements}

We gratefully acknowledge the support of the Office of Naval Research grant N0014-20-1-2253.

%
%
%

\bibliographystyle{unsrt}
\bibliography{citations}

\end{multicols*}
\end{document}